\documentclass[
    ,final            % use final for the camera ready runs
  ]
  {aipproc}

\layoutstyle{8x11double}
\def\bi{\begin{itemize}}
\def\ei{\end{itemize}}
\def\bt{\begin{table}}
\def\et{\end{table}}
\def\N0{\widetilde{\chi}^0}
\def\Dt{\widetilde{\Delta}}
\def\Dp{\widetilde{\Delta}^{++}}
\def\Dm{\widetilde{\Delta}^{--}}
\def\Dt{\widetilde{\Delta}}
\def\Cp{\widetilde{\chi}^+}
\def\Cm{\widetilde{\chi}^-}
\def\Cpm{\widetilde{\chi}^\pm}

\def\sel{\widetilde{e}_{L}}
\def\ser{\widetilde{e}_{R}}
\def\sml{\widetilde{\mu}_{L}}
\def\smr{\widetilde{\mu}_{R}}
\def\stl{\widetilde{\tau}_{1}}
\def\str{\widetilde{\tau}_{2}}
\def\slash {\!\!\!\!/}

\begin{document}

\title{Signatures for doubly-charged Higgsinos at colliders
\footnote{Based on the talk presented at SUSY 2008 conference in Seoul, Korea.}
}

\classification{12.60.Jv, 12.60.Fr}
\keywords      {left-right supersymmetry, doubly-charged Higgsinos.}

\author{D. A. Demir}{
  address={Department of Physics, Izmir Institute of Technology, IZTECH,
TR35430 Izmir, Turkey.}
 ,altaddress={Deutsches Elektronen - Synchrotron, DESY, D-22603 Hamburg,
Germany.} % additional visiting address
}

\author{M. Frank}{
  address={Department of Physics, Concordia University,
7141 Sherbrooke Street West, Montreal, Quebec, CANADA H4B 1R6.}
}
\author{K. Huitu}{
  address={Department of Physics, University of Helsinki and Helsinki 
Institute of Physics, P.O. Box 64, FIN-00014, Helsink, Finland}
}

\author{\underline {S. K. Rai}}{
  address={Department of Physics, University of Helsinki and Helsinki 
Institute of Physics, P.O. Box 64, FIN-00014, Helsink, Finland}
}

\author{I. Turan}{
  address={Department of Physics, Concordia University,
7141 Sherbrooke Street West, Montreal, Quebec, CANADA H4B 1R6.}
}

\begin{abstract}
Several supersymmetric models with extended gauge structures predict light 
doubly-charged Higgsinos. Their distinctive signature at the large hadron 
collider is highlighted by studying their production and decay characteristics. 
\end{abstract}

\maketitle

%%%%%%%%%%%%%%%%%%%%%%%%%%%%%%%%%%%%%%%%%%%%
%% MAINMATTER
%%%%%%%%%%%%%%%%%%%%%%%%%%%%%%%%%%%%%%%%%%%%

\section{Introduction}

Supersymmetry (SUSY), in the form of the minimal supersymmetric standard model
(MSSM) is the most popular scenario of physics beyond the standard model (SM)
and resolves the gauge hierarchy problem very elegantly. 
A general, although not universally present, feature of SUSY, is that, 
if R-parity $R=(-1)^{(3B+L+2S)}$ (with $B$, $L$ and $S$ being baryon, lepton 
and spin quantum numbers, respectively) is conserved, the absolute stability 
of the lightest supersymmetric partner (LSP) is guaranteed, making
it a viable candidate for cold dark matter in the universe. 
However, to generate neutrino masses, one must either invoke R-parity 
violation and abandon the 
absolute stability of LSP or add right-handed neutrinos and introduce the seesaw
mechanism. Supersymmetric grand unified theories (SUSY GUTs) which contain 
left-right supersymmetry (LRSUSY) resolve both problems naturally 
\cite{history,Demir:2006ef}. The LRSUSY gauge theory is based on the product 
group $SU(3)_C \times SU(2)_L \times SU(2)_R \times U(1)_{B-L}$. 
LRSUSY models are attractive for many reasons. They disallow explicit R-parity
breaking in the Lagrangian; they provide a natural mechanism for
generating neutrino masses; and they provide a solution to the strong and
weak CP problem in MSSM \cite{Mohapatra:1995xd}. Neutrino masses are
induced by the see-saw mechanism through the introduction of Higgs triplet
fields which transform as the adjoint of the $SU(2)_R$ group and have
quantum numbers $B-L=\pm 2$.
While the Higgs triplet bosons are present in the
non-supersymmetric version of the theory, their fermionic partners, the
Higgsinos, are specific to the supersymmetric version. It has been shown
that, if the scale for left-right symmetry breaking is chosen so that the
light neutrinos have the experimentally expected masses, these Higgsinos
can be light, with masses in the range of ${\cal O} (100)$ GeV
\cite{Chacko:1997cm}. Such particles could be produced in abundance 
and thus give definite signs of left-right symmetry at
future colliders like the LHC and linear $e^+e^-$ colliders. 
For a more detailed information about the model see, for instance 
\cite{history, Francis:1990pi}. 

As the underlying 
theory to model "new physics" at the TeV scale bring about new particles and 
interaction schemes, experiments at the Large Hadron Collider (LHC) will be 
probing these new particles as well as new interactions. 
The models of new physics will be distinguished by certain characteristic 
signatures in regard to their lepton and jet spectra in the final state. In 
this talk we present signatures specific to the LRSUSY model at the LHC.
We consider the pair production and single production  of the doubly-charged 
Higgsinos at LHC and analyze the signals resulting from their decays. We refer 
the readers to \cite{rai1,rai2} for a more detailed account on the analysis.

\begin{table}
\begin{tabular}{lrrr}
\hline
    \tablehead{1}{l}{b}{fields\tablenote{all masses in GeV}}     
  & \tablehead{1}{c}{b}{SPA}
  & \tablehead{1}{c}{b}{SPB}
  & \tablehead{1}{c}{b}{SPC} \\
%\hline
%     &$\tan\beta=5,M_{B-L}= 25$ GeV
%     &$\tan\beta=5,M_{B-L}= 100$ GeV
%     &$\tan\beta=5,M_{B-L}= 0$ GeV\\
%     & $M_L=M_R=250$ GeV
%     & $M_L=M_R=500$ GeV
%     & $M_L=M_R=500$ GeV \\
%     &$v_{\Delta_R}=3000$ GeV,$v_{\delta_R}=1000$ GeV
%     &$v_{\Delta_R}=2500$ GeV,$v_{\delta_R}=1500$ GeV
%     &$v_{\Delta_R}=2500$ GeV,$v_{\delta_R}=1500$ GeV\\
%     &$\mu_1 = 1000$ GeV,$\mu_3 =300$ GeV
%     &$\mu_1 = 500$ GeV,$\mu_3 =500$ GeV
%     &$\mu_1 = 500$ GeV,$\mu_3 =300$ GeV \\
\hline
$\N0_i~(i=1,3)$ & 89.9, 180.6, 250.9 & 212.9,441.2,458.5 & 142.5, 265.6, 300.0\\
$\Cpm_i~(i=1,3)$& 250.9,300.0,953.9 & 459.4, 500.0,500.0 & 300.0, 459.3, 500.0\\
$M_{\Dt}$       & 300 & 500 & 300 \\
$W_R, Z_R$    & 2090.4, 3508.5    & 1927.2, 3234.8 & 1927.2, 3234.8 \\ 
              & {\bf S2} ~~~~~~~~~~~~~~~~~~ {\bf S3}
              & {\bf S2} ~~~~~~~~~~~~~~~~~~ {\bf S3}
              & {\bf S2} ~~~~~~~~~~~~~~~~~~ {\bf S3}    \\
$\sel,\ser$     & (156.9,155.6), (402,402) & (254.2,253.4),
(552,552) & (214.9,214.0), (402.6,402.2)\\
$\sml,\smr$     & (156.9,155.6), (402,402) & (254.2,253.4),
(552,552) & (214.9,214.0), (402.6,402.2)\\
$\stl,\str$     & (155.4,159.9), (401,406) & (252.5,257.9),
(550,556) & (212.8,216.2), (401.5,403.3)\\\hline
\end{tabular}
\caption{The low lying mass spectrum in the model, 
defining the sample points {\bf SPA}, {\bf SPB} and {\bf SPC}. In each case,
{\bf S2} and {\bf S3} designate parameter values which allow for {\bf 2}-body
and {\bf 3}-body decays of doubly-charged Higgsinos, respectively.}
\label{susyin}
\end{table}
\subsection{Production and decay of doubly-charged Higgsinos}
The pair-production process at the LHC is
$p\, p \longrightarrow \Dp\, \Dm $ 
which proceeds with $s$-channel $\gamma$ and $Z_{L,R}$ exchanges, and
the associated production mode
$p\, p \longrightarrow \Cp_1\, \Dm $ 
which rests on $s$-channel $W_{L,R}$ exchanges. Both processes are
generated by quark--anti-quark annihilation at the parton level.
These doubly-- and singly--charged fermions  subsequently
decay via a chain of cascades until the lightest neutralino $\chi_1^0$
is reached. In general, the two-body
decays of doubly-charged Higgsinos are given by
\bi
\item $\Dm \longrightarrow \widetilde{\ell}^- ~\ell^-, \Delta^{--} ~\N0_i,
\Cm_i ~\Delta^{-}, \Cm_i ~W^{-}$
\ei
whose decay products further cascade into lower-mass
daughter particles of which leptons are of particular
interest. 
%%%%%%%%%%%%%%%%%%%%%%%%%%%%%%
\begin{figure}
  \includegraphics[height=.225\textheight]{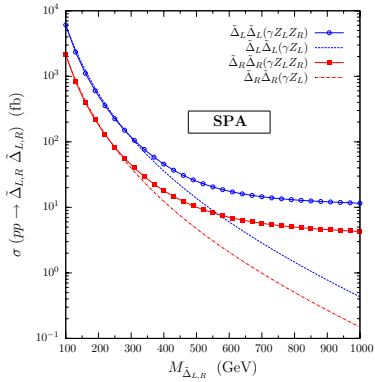} \label{paircs}
  \caption{The pair production cross sections for doubly-charged Higgsinos
in LRSUSY at the LHC.}
\end{figure}
%%%%%%%%%%%%%%%%%%%%%%%%%%%%%%

We choose three sample points in the LRSUSY parameter space, as tabulated in
Table \ref{susyin}. A quick look at the mass spectrum for the
sparticles suggest that the chargino states are also heavier than or 
comparable to the
doubly-charged Higgsinos, and hence, the favorable decay channel for
$\Dt$ is $\Dm \longrightarrow \widetilde{\ell}^- ~ \ell^-$ ({\bf S2}), provided
that $m_{\tilde{l}} < M_{\Dm}$. For relatively light Higgsinos, one can, in
principle, have $m_{\tilde{l}} > M_{\Dm}$ in which case the
only allowed decay mode for the doubly-charged
Higgsinos would be the 3-body decays, which would proceed dominantly through off-shell sleptons:  $\Dm \to
\widetilde{\ell}^{\star\, -} ~ \ell^- \to \ell^- \ell^- \N0_1$ ({\bf S3}). 

For the benchmark point in Table~\ref{susyin}, the doubly-charged Higgsinos
assume the following  2-- and 3--body decay branchings:
% ------------------------------------------------------------------
\begin{eqnarray}
\label{eq:decays}
 BR(\Dm_{L/R} \to \tilde{\ell}_{iL/iR}^-\ell_i^-) &\simeq& \frac{1}{3},
~~~m_{\tilde l_{i}}<M_{\tilde\Delta^{--}} \nonumber\\
 BR(\tilde{\ell}_{iL/iR}^- \to \ell_i^- \N0_1) &=& 1, \\
 BR(\Dm_{L/R} \to \ell_i^- \ell_i^-  \N0_1) &\simeq& \frac{1}{3},
~~~m_{\tilde l_{i}}>M_{\tilde\Delta^{--}} \nonumber
\end{eqnarray}
% ------------------------------------------------------------------
where $i=e,\mu,\tau$. 
%%%%%%%%%%%%%%%%%%%%%%%%%%%%%%
\begin{figure}
  \includegraphics[height=.225\textheight]{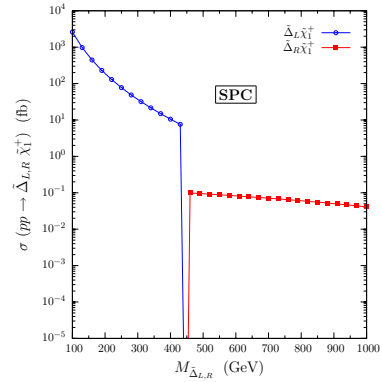} \label{assocs}
  \caption{The cross sections for associated productions of
$\Dt_{L,R}$ and $\tilde{\chi}_1^{\pm}$ in the LRSUSY model at LHC.}
\end{figure}
%%%%%%%%%%%%%%%%%%%%%%%%%%%%%%
%\subsection{Pair-production and single production of doubly-charged Higgsinos}

The pair-production and associated production cross sections are shown in
Figs \ref{paircs},\ref{assocs}.  
The doubly-charged Higgsinos decay according to \eqref{eq:decays} into
two same-sign same-flavor (SSSF) leptons and the lightest neutralino $\N0_1$, 
the LSP. This decay pattern, when the doubly charged states are produced in 
pair, gives rise to final states involving four isolated leptons of the form 
%%%%%%%%%%%%%%%%%%%%%
\begin{equation}
p p \longrightarrow \Dp \Dm \longrightarrow  \left(\ell^+_i \ell^+_i\right) +
\left(\ell^-_j \ell^-_j\right) + E\slash_T\,,
\label{eq:pairprod}
\end{equation}
%%%%%%%%%%%%%%%%%%%%%
where $\ell_i,\ell_j=e,\mu,\tau$. Similarly, in the case of associated 
production one gets
%%%%%%%%%%%%%%%%%%%%%
\begin{equation}
p\, p \longrightarrow \Dm\, \Cp_1 \longrightarrow
 \left(\ell_i^-\ell_i^-\right) +  \ell_j^+ + E\slash_T , \nonumber
\label{eq:singprod}
\end{equation}
%%%%%%%%%%%%%%%%%%%%%

%\paragraph{$4\ell+E\slash_T$}

%%%%%%%%%%%%%%%%%%%%%%%%%%%%%%
\begin{figure}
  \includegraphics[height=.215\textheight]{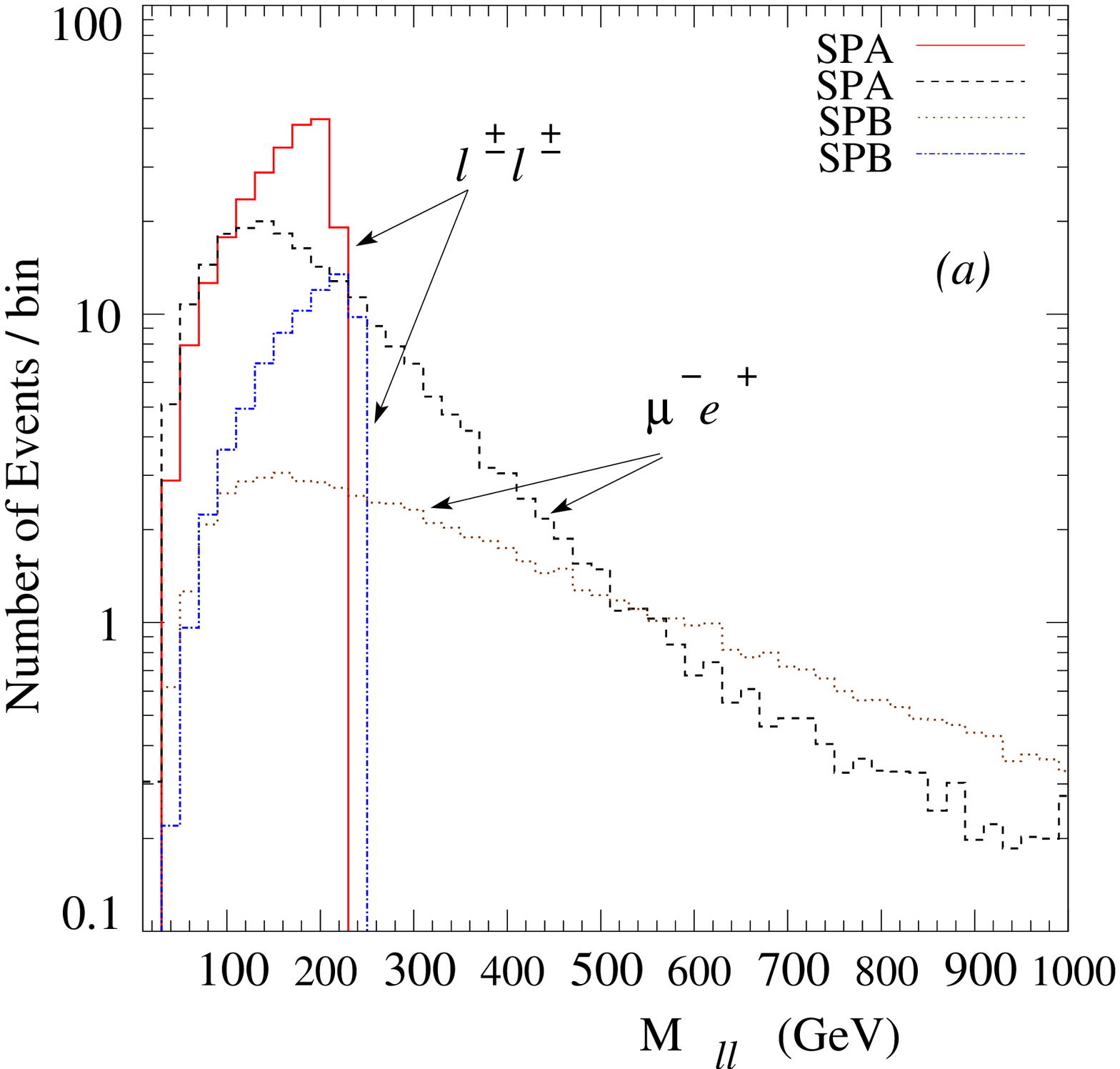} 
  \includegraphics[height=.215\textheight]{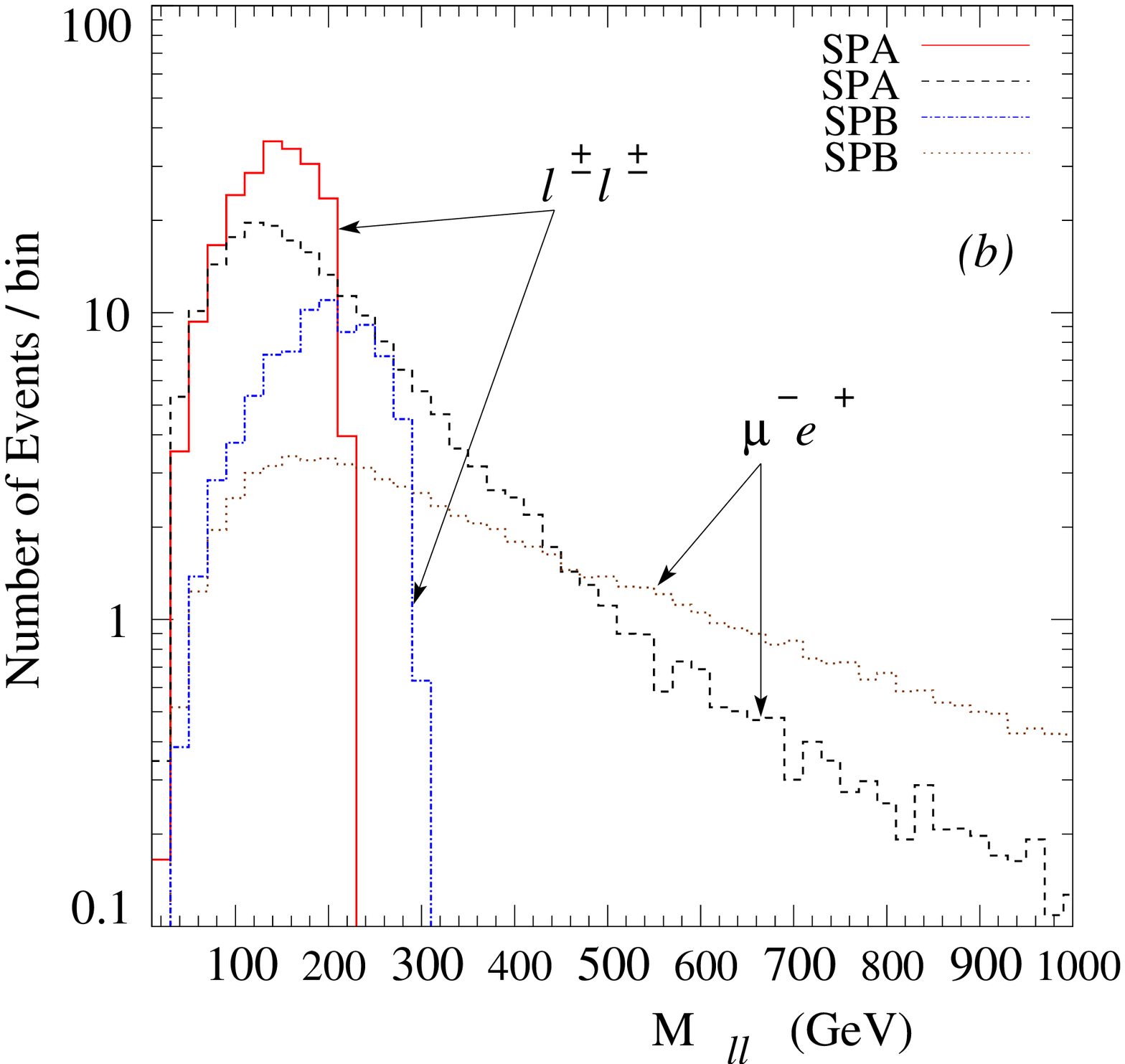} 
  \caption{Binwise invariant mass distribution of lepton pairs with 
integrated luminosity of $\int{\cal L} dt = 30 fb^{-1}$. The panel (a) represents  the 2-body ({\bf S2}) case, and panel (b) does the 3-body ({\bf S3}) case.}
\label{MM4l}
\end{figure}
%%%%%%%%%%%%%%%%%%%%%%%%%%%%%%
The $4\ell+E\slash_T$ signal receives contributions from the pair-production of
both chiral states of the doubly-charged Higgsino so we add up
their individual contributions to obtain the total number
of events. This yields a rather clean and robust $4\ell+$ missing $p_T$
signal at the LHC with highly suppressed SM background.
We impose the following kinematic cuts on our final states:
%%%%%%%%%%%%%%%%%%
\begin{itemize}
\item  The charged leptons in the final state satisfy $|\eta_{\ell}|<2.5$,
and have a minimum transverse momentum $p_T > 25$ GeV.
Each pair of oppositely-charged leptons of same flavor have at least
$10\ {\rm GeV}$ invariant mass.
\item To ensure proper resolution between the final state leptons we demand
$\Delta R_{\ell\ell}>0.4$ for each pair of leptons.
\item The missing transverse energy must be $E\slash_T > 50$ GeV.
\end{itemize}
%%%%%%%%%%%%%%%%%%%
The total cross section for the $2\mu^- + 2e^+ +E\slash_T$ signal for 
{\bf SPA} point is 7.71 fb ({\bf S2}) and 7.02 fb ({\bf S3}) while 
for {\bf SPB} it is 2.43 fb ({\bf S2}) and 2.66 fb ({\bf S3}). 
In the case where the doubly-charged Higgsino is produced in association with 
a chargino, we have used the {\bf SPC} point, and the corresponding cross sections 
for $2\mu^- + e^+ + E\slash_T$ are 2.24 fb ({\bf S2}) and 
2.03 fb ({\bf S3}), respectively. 
As in pair-production, the $\Dm$ decays again into a pair of SSSF leptons
and an LSP, while the chargino decays with almost $100\%$ branching ratio 
to a neutrino and slepton for {\bf SPC}. 
In Fig. \ref{MM4l} we show the binwise invariant mass distribution of leptons 
pairs for both {\bf SPA,SPB} points and both cases of {\bf S2} and {\bf S3}.
These plots manifestly show differences between the SSSF and 
opposite-sign-different-flavor (OSDF)
lepton pairs in regard to their invariant mass distributions. Indeed,
the SSSF lepton pairs exhibit a sharp kinematic edge according to the 
expression given by 
\begin{equation}
M_{\ell^\pm\ell^\pm}^{max} =
\sqrt{M_{\Dt}^2 + M_{\N0_1}^2 - 2 M_{\Dt} E_{\N0_1}}\,\,,
\label{eq:invmass}
\end{equation}
where $E_{\N0_1}$ is the energy of the LSP, in their $M_{\ell\ell}$
distributions whereas the OSDF lepton pairs do not. These act as very 
effective discriminants when compared to similar signals in MSSM. 
The edge in the SSSF dilepton invariant mass distribution yields a clear hint
of a $\Delta L=2$ interaction and a doubly-charged field in the underlying model
of `new physics'. A very similar feature is expected for {\bf SPC} point and the
$3\ell+E\slash_T$ signal. The invariant mass distribution and also the
distribution in $\Delta R$ of a pair SSSF and OSDF leptons, prove to be 
a very effective discriminant when compared with similar signals coming
from other new physics scenarios. We refer the readers to \cite{rai2} for 
further details.

%\begin{description}
%\item[Infandum]
%\end{description}

%\begin{theacknowledgments}
%\end{theacknowledgments}




%%%%%%%%%%%%%%%%%%%%%%%%%%%%%%%%%%%%%%%%%%%
%% The following lines show an example how to produce a bibliography
%% without the help of the BibTeX program. This could be used instead
%% of the above.
%%%%%%%%%%%%%%%%%%%%%%%%%%%%%%%%%%%%%%%%%%%

\end{document}